\documentclass[12pt,final]{iopart}
\usepackage{latexsym}
\usepackage{graphics}
\usepackage{graphicx}
\usepackage{amssymb}
\usepackage[latin1]{inputenc}
\newcommand{\BSCCO}{Pb-BSCCO}
\newcommand{\BiSCCO}{Bi$_{1.8}$Pb$_{0.26}$Sr$_2$Ca$_2$Cu$_3$O$_{10+x}$}
\begin{document}

\title[Pb-BSCCO Coaxial cavity resonator]{Tunable coaxial cavity resonator for linear and nonlinear microwave characterisation of superconducting wires}

\author{A. Agliolo Gallitto, G. Bonsignore, M. Li Vigni and A. Maccarone}
\address{CNISM and Dipartimento di Fisica, Università di Palermo, Via Archirafi 36, I-90123 Palermo (Italy)}

\begin{abstract}
We discuss experimental results obtained using a tunable cylindrical coaxial cavity constituted by an outer Cu cylinder and an inner \BiSCCO\ wire. We have used this device for investigating the microwave response of the superconducting wire, both in the linear and nonlinear regimes. In particular, by tuning the different modes of the cavity to make them resonant at exactly harmonic frequencies, we have detected the power emitted by the superconducting inner wire at the second- and third-harmonic frequency of the driving field. The results obtained in the nonlinear regime, whether for the microwave surface impedance or the harmonic emission, are qualitatively accounted for considering intergrain fluxon dynamics. The use of this kind of device can be of strong interest to investigate and characterise wires of large dimensions to be used for implementing superconducting-based microwave devices.
\end{abstract}


\maketitle

\section{Introduction}
Superconducting materials, due to their low microwave (\emph{mw}) surface impedance, are preferred to normal metals for assembling \emph{mw} devices, such as filters, antennas, resonators, etc.~\cite{Lancaster,hein,Gallop,padam}. Investigation of the \emph{mw} response of superconductors (SC), besides its importance for basic physics, gives information on specific properties of the investigated sample, allowing to recognise the materials suitable for designing devices operating at \emph{mw} frequencies~\cite{padam,Gallop}. Since the discovery of high-$T_c$ superconductors (HTSC), several studies have been carried out to exploit the potential of these SC for implementing \emph{mw} devices; comprehensive reviews on \emph{mw} applications of HTSC have been reported in Refs.~\cite{Lancaster,hein,Gallop}.

Actually, the main factor limiting the use of HTSC in passive \emph{mw} devices is the occurrence of nonlinear effects at high input power levels. Nonlinear effects manifest themselves in  the power dependence of the surface impedance~\cite{samoilova,golo_NL,nguyen,velichko,Oates_libro}, intermodulation product~\cite{samoilova,velichko,Oates_libro,oates} and harmonic generation~\cite{samoilova,Noi_FisicaC161,noi_review,Noi-Metamat}, which worsen the performance of passive \emph{mw} devices; so, it is of importance to estimate the power and/or field values at which nonlinearity comes into play, as well as to understand its physical origin.

One of the $mw$ devices mostly used in both applications and basic-physics studies of SC is the resonant cavity. On the technological point of view, superconducting resonant cavities can be conveniently used in all the systems requiring high selectivity in the signal frequency. On the other hand, cavity-perturbation technique allows measuring the $mw$ surface impedance, $Z_s=R_s + i X_s$, of superconducting materials, which is the most important property characterizing the \emph{mw} response~\cite{Cavity_pert,Owliaei,Noi-BKBO,Noi-CrSt}. With the advance of the modern techniques for film deposition, planar transmission line filters or strip-line resonators of small size can be fabricated and used in many applications. Despite the availability of these smaller sized filters, bulk- or coated-cavity filters provide higher quality factor and reduced nonlinear effects. This is so because small cross-section areas for current flow  in films lead to high current densities, even at relatively low input power levels, enhancing nonlinear effects~\cite{Oates_libro,Dam_Scalapino}. Because of these properties, superconducting films have been recommended for implementing \emph{mw} power limiters~\cite{Booth}. Bulk-cavity filters are preferred in all the applications in which miniaturisation is not important~\cite{Pandit}, as for example high-frequency wireless communication systems, satellite transmission systems, radars, high-power transmission lines.

Characterisation at \emph{mw} frequencies of SC is generally performed by locating a small sample in resonant cavities operating at fixed frequencies. To investigate the surface impedance, one measures the variations of the resonant frequency and the quality factor of the cavity induced by the sample~\cite{Cavity_pert}. Intermodulation product is investigated by applying two \emph{mw} signals, whose frequencies are within the bandwidth of the cavity, and detecting third order intermodulation frequencies by a spectrum analyzer~\cite{Oates_libro}. Harmonic generation is investigated by using a nonlinear \emph{mw} spectrometer, whose basic element is the bimodal cavity, resonating at two different frequencies ($\omega$ and $n\omega$)~\cite{Noi-Metamat,leviev}; different cavities have to be used for detecting the $n$-order harmonic signal. So, for a complete \emph{mw} characterisation different setups and cavities are necessary.

Coaxial cavity resonator with superconducting inner conductor has been proposed to conveniently measure the frequency dependence of the \emph{mw} surface resistance, $R_s$~\cite{Lancaster}. Prototypes of coaxial resonators have been built using normal metal as outer conductor and HTSC as inner conductor and used for measuring the \emph{mw} surface resistance of the inner SC~\cite{delayen,BiSCCO_R(H),YBCO_Z(f)}. The use of such kind of resonator is particularly suitable to characterise samples having dimensions too large for using the cavity-perturbation method. The inner conductor can be constituted by bulk or coated slabs and/or wires. The length can be dimensioned to investigate different frequency ranges. High sensitivity can be achieved by using thin central wires or SC materials for both the inner wire and the outer cylinder. However, for investigating the \emph{mw} properties of SC as a function of a DC magnetic field, the outer cylinder should be made of normal metal. In this paper, we discuss the potential of such type of device for a complete investigation of SC in both the linear and nonlinear regimes. We have build a tunable cylindrical coaxial cavity constituted by an outer Cu cylinder and an inner \BiSCCO\ wire and have used it for different types of measurements. In the linear regime, we have measured $R_s$ of the inner wire as a function of the frequency, at zero DC magnetic field. In the nonlinear regimes, we have measured the power dependence of the surface impedance and the signals emitted by the SC at the second- and third-harmonic frequencies of the driving field. In order to detect the harmonic signals, the different modes of the cavity have been tuned to make them resonant at exactly harmonic frequencies. This has been done by inserting two metallic screws in the outer conductor, which modify the impedance of the whole cavity and slightly change the resonant frequencies. This trick allows one to measure all the nonlinear properties of the investigated material in the same excitation conditions.

\section{Coaxial cavity resonator: theoretical aspects}\label{teoria}
A coaxial cavity resonator consists of an outer conductor tube containing an inner wire. An extensive discussion of this \emph{mw} device is reported in~\cite{Lancaster}; here, we discuss only the essential concepts useful for understanding the present work. Coaxial resonator supports TEM, TM and TE modes; however, the design of our cavity is such that, in the investigated frequency range ($1 \div 13$ GHz), the only detectable modes are the TEM ones, corresponding to standing waves in which an integer number of half-wavelength nearly matches with the length of the inner conductor. In these modes, electric-field lines are radial and magnetic-field lines wind around the central conductor; in an open-circuit-end resonator, as it is our cavity, different TEM modes can be fed, with the condition that at both the inner-conductor ends the electric field is maximum and magnetic field is zero. The quality factor of a coaxial cavity with open-circuit ends, operating in the TEM mode, is given by
\begin{equation}\label{Qc}
   Q_c = \frac{\omega \mu_0 \ln (b/a)}{\frac {R_{sa}}{a}+\frac {R_{sb}}{b}}\,,
\end{equation}
where $a$ is the radius of the inner conductor, $b$ the inner radius of the outer conductor, $R_{sa}$ and $R_{sb}$ are the surface resistances of the inner and outer conductors, respectively, and $\omega$ is the angular frequency of the considered mode.

Eq.~(\ref{Qc}) includes only the energy losses occurring in the walls of the outer and inner conductors; if the cavity is filled with a dielectric medium, the unloaded quality factor of the resonator, $Q_U$, is affected by conductor losses and dielectric losses and it is given by
\begin{equation}\label{Q_U}
    \frac{1}{Q_U}=\frac{1}{Q_c}+\frac{1}{Q_d}\,,
\end{equation}
where $Q_d$ takes into account the dielectric losses; when the dielectric completely fills the cavity, $1/Q_d=\tan \delta$.\\
Moreover, since the cavity is generally coupled to an external circuit through an excitation and a detection line, additional energy losses out of the coupling ports occur and the measured (loaded) quality factor, denoted as $Q_L$, differs from $Q_U$. If the cavity is weakly coupled to the external circuit, only small corrections have to be done to calculate $Q_U$; in any case, this correction can be easily calculated considering the coupling coefficients (see Chap. 4 of Ref.~\cite{Lancaster}).

From Eqs. (\ref{Qc}) and (\ref{Q_U}), the \emph{mw} surface resistance of the inner conductor can be written as
\begin{equation}\label{RsSC}
    R_{sa}= \Gamma \left[\frac{1}{Q_U}- \tan \delta \right] - \frac{a}{b} R_{sb}\,,
\end{equation}
where $\Gamma$ is the geometry factor, given by
\begin{equation*}
  \Gamma = a \omega \mu_0 \ln (b/a)\,.
\end{equation*}
From Eq.~(\ref{RsSC}), one can see that to determine $R_{sa}$ it is necessary to know $R_{sb}$ and $\tan \delta$.

\section{The tunable coaxial cavity}\label{cavity}
The \emph{mw} coaxial cavity consists of a copper outer tube, of length $h= 43.8$~mm and inner diameter $2b=8.1$~mm,  with the dielectric insulator PTFE supporting an inner \BiSCCO\ (\BSCCO) superconducting wire, of length $l=40$~mm and diameter $2a=3$~mm. The \BSCCO\ wire has been produced by Can Superconductors s.r.o.; material characterisation can be found in the web site of the company. The cavity is used in the transmission mode; it is coupled to the external excitation and detection lines by capacitive gap at both ends of the resonator. The gap between the superconducting wire and the feed wire is adjusted to give weak coupling, so that measurements of $Q_L$ require only small corrections to calculate $Q_U$. Two small plates of PTFE are also located at the ends of the \BSCCO\ wire, in order to prevent intermittent electrical contact. A schematic draw of the cavity is shown in Fig.~\ref{cavity}.
\begin{figure}[h]
  \centering
  \includegraphics[width=8 cm]{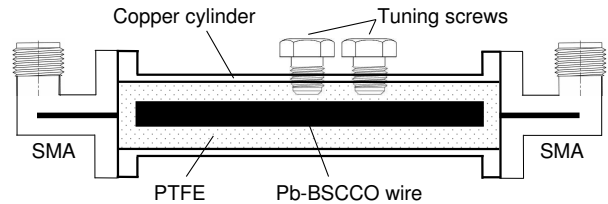}
  \caption{Schematic diagram of the coaxial cavity.}
  \label{cavity}
\end{figure}

Because of the capacitive effects at the gaps, the electromagnetic field extends slightly beyond the ends of the wire. This causes small differences of the resonant frequencies from the expected ones; for the same reason, the resonant frequencies of the higher-frequency modes are not exactly harmonics of the fundamental one. In order to tune the first three modes of the cavity and make them resonant at exactly harmonic frequencies, we have made two holes in the outer conductor for inserting two metallic screws that, modifying the impedance of the whole cavity, slightly change the resonant frequencies. As shown in Fig.~1, a screw is located in proximity of the middle length of the outer conductor; by regulating the penetration of this screw into the cavity, we can tune the first two modes within the frequency bandwidth. A second screw, located at about one third of the length, allows us to tune the first and third modes too. This procedure is done to feed the cavity in the fundamental mode and detect the second-harmonic (SH) and third-harmonic (TH) signals emitted by the SC wire.

The cavity has been used to perform several kinds of measurements. $R_s$ of the \BSCCO\ wire has been measured, at fixed temperatures (below $T_c$), as a function of the frequency and input power level. Furthermore, we have investigated the signals emitted by the wire at the SH and TH frequencies of the driving field. In order to make measurements as a function of the external magnetic field, the cavity is placed between the poles of an electromagnet that generates DC magnetic fields up to $\mu_0 H_0 \approx 1~$T; $\mathbf{H_0}$ can be rotated in the plane containing the cylinder axis to change its orientation with respect to the \emph{mw} magnetic field. Two additional coils, independently fed, allow compensating the residual magnetic field and making measurements at low magnetic fields. 

\section{Microwave surface resistance}
\subsection{Measurement technique}
As one can see from Eq.~(\ref{RsSC}), in order to determine $R_{sa}$ of the \BSCCO\ wire, we need to know $Q_U$, the \emph{mw} surface resistance of the outer conductor, $R_{sb}$, and $\tan \delta$. At low input power ($P_{in} \lesssim 0$~dBm), $Q_L$ has been measured using an HP8719D network analyzer (NA), operating in the frequency range 50~MHz - 13.5~GHz, and detecting the frequency response of the transmitted wave. $Q_U$ has been determined by taking into account the coupling coefficients, $\beta_1$ and $\beta_2$, for both the coupling lines; these coefficients can be calculated by directly measuring the reflected power at each line, as described in Ref.~\cite{Lancaster}, Chap.~4. $Q_U$ is given by $Q_U = (1+\beta_1+\beta_2)Q_L$.

To determine the frequency dependence of $R_{sa}$, we have previously calibrated the cavity using an inner copper wire having the same dimension of the \BSCCO\ wire. At fixed frequency, and in absence of PTFE, we have determine the surface resistance of the copper, from Eq.~(\ref{Qc}) letting $R_{sa}=R_{sb}$. Successively, we have investigated the frequency response of the cavity with the dielectric insulator PTFE supporting the inner copper wire, and measured the quality factor and the central frequency in the different resonant modes. By assuming the $\sqrt{\omega}$ law for the frequency dependence of $R_{sb}$, we have determined $\tan \delta$ as a function of the frequency. We have obtained for the loss tangent values of the order of $10^{-4}$, slowly increasing with the frequency. Once $Q_U(f)$, $R_{sb}(f)$ and $\tan \delta(f)$ are known, from Eq.~(\ref{RsSC}) we obtain the frequency dependence of the \emph{mw} surface resistance of the SC wire in the frequency range in which the different resonant modes fall.

Measurements of \emph{mw} surface impedance as a function of the maximum value of the \emph{mw} magnetic field, $H_{mw}$, have been performed varying the input power levels and using two different \emph{mw} sources, for the fundamental mode and the higher frequency modes. For the fundamental mode ($f=\omega /2\pi \approx 2.525$~GHz), the continuous wave generated by the NA is modulated to obtain a train of \emph{mw} pulses, with pulse width $\approx 10~\mu$s and pulse repetition rate 100~Hz. The pulsed signal is amplified up to a peak power level of $\approx$~44~dBm and driven into the cavity through the excitation line. The transmitted signal is detected by a solid-state diode. The resonance curve of the cavity is reproduced by sweeping the continuous-wave frequency of the NA. By a Lorentzian fit of the resonance curve, we extract the central frequency and the unloaded quality factor of the cavity. From $Q_U$, we have determined the surface resistance of the sample as above discussed; from the central frequency, we have determined the surface reactance, $X_s$. The surface reactance can be determined measuring the shift of the resonance frequency of the cavity, apart from an additive constant $X_0$. To obtain $X_s$ in absolute units, it is necessary to know $X_0$, which can be determined by imposing the condition of normal skin effect at $T = T_c$. Since in the normal state the resonance curve is broadened and deformed, we will report only the variation of $X_s$ with respect to that obtained at the lowest \emph{mw} magnetic field, which is given by
\begin{equation}
    \Delta X_s \equiv X_s(H_{mw})-X_s(0)= -2~\Gamma \frac {f(H_{mw})-f(0)}{f(0)}\,.
\end{equation}
The values of the $H_{mw}$ have been calculated considering the input power, the unload quality factor of the cavity and the coupling coefficients in the different modes, following Chap.~3 of Ref.~\cite{Lancaster}.

For the higher frequency modes, since we do not dispose of \emph{mw} amplifier at frequencies higher than 4~GHz, we have used a \emph{mw} sweeper that generates a continuous wave up to power levels of $\approx$~20 dBm. Also in this case, the resonance curve of the cavity is reproduced by sweeping the continuous-wave frequency.

\subsection{Results and discussion}
\begin{figure}[h!]
\centering
\includegraphics[width=8cm]{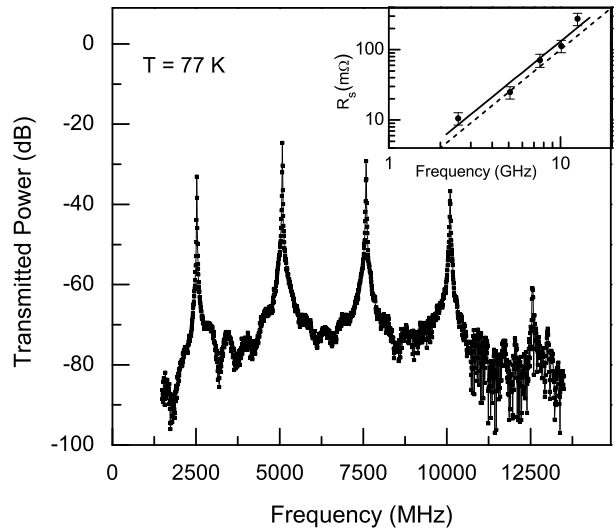}
\caption{Transmitted power over input power through the coaxial cavity as function of the frequency. Inset: frequency dependence of $R_s$ of the \BSCCO\ wire; the continuous line is the best-fit curve of data, the dashed one represents the $f^2$ law.}
\label{spettro}
\end{figure}

Figure~\ref{spettro} shows the \emph{mw} response of the coaxial cavity resonator with the \BSCCO\ wire in the frequency range $1 \div 13$~GHz, at $T = 77$~K and $P_{in} \approx -1$~dBm. The inset shows the frequency dependence of $R_s$ of the \BSCCO\ wire, at the same temperature and input power level. The values of $R_s$ are consistent with those reported in the literature for HTSC ceramic samples~\cite{BiSCCO_R(H),YBCO_Z(f),Piel}. The continuous and the dashed lines in the inset show the best-fit curve of data and the $f^2$ law, respectively.

The $f^2$ dependence of $R_s$ is expected in the London two-fluid model as well as in the BCS theory when the imaginary component of the complex conductivity is much larger than the real one.  It has been experimentally shown that it is valid also in HTSC single crystals and high quality thin films. For bulk HTSC materials, it has been observed $R_s \propto f^n$ with $n$ ranging from 1 to 2, depending on the investigated sample~\cite{delayen,YBCO_Z(f),AlfordIEEE,AlfordSUST}. Deviations from the $f^2$ law have been ascribed to normal material inclusion at the grain boundaries, poor surface quality, wide variation of crystal-axis orientation of the grains, etc.~\cite{AlfordIEEE,AlfordSUST}. On the other hand, models developed to describe the electromagnetic response of weakly coupled grains have shown that different $f$ dependencies are expected in the different field and frequency ranges~\cite{halbritter}. In particular, a $f^2$ law is expected at \emph{mw} magnetic fields lower than the Josephson lower critical field, $H_{c1j}$. To our knowledge, results on bulk BSCCO samples are not reported in the literature, most investigations concern YBCO samples. The data of Fig.~\ref{spettro} have been obtained with $H_{mw} < 10~\mu$T; so, the nearly quadratic frequency dependence of $R_s$ we obtain can be ascribed to the fact that $H_{mw} < H_{c1j}$.

\begin{figure}[h]
\centering
\includegraphics[width=8cm]{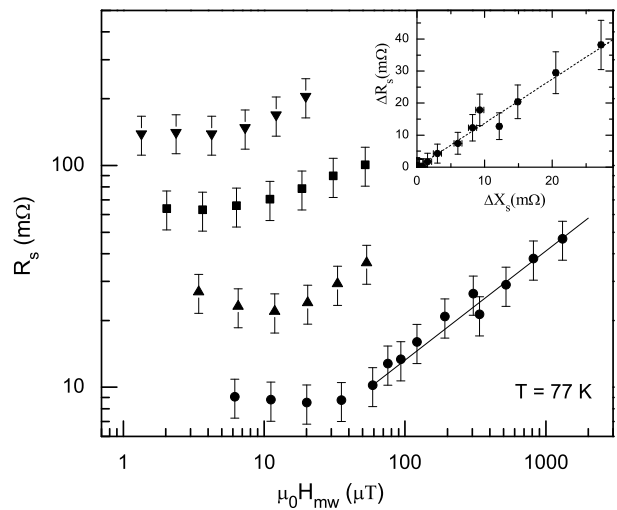}
\caption{$R_s$ of the \BSCCO\ wire as a function of the maximum value of the \emph{mw} magnetic field, at the resonant frequencies of the different TEM modes: $f\approx 2.5$~GHz ($\bullet$); $f\approx 5.1$~GHz ($\blacktriangle$); $f\approx 7.6$~GHz ($_\blacksquare$); $f\approx 10.1$~GHz ($\blacktriangledown$). The uncertainties in $R_s$ ($\approx 20 \%$) are essentially due to the determination of $\tan \delta$ of PTFE. The inset shows the variations of $R_s$ as a function of the variations of $X_s$, obtained at different values of $H_{mw}$, for $f\approx 2.5$~GHz.}
\label{Rs(P)}
\end{figure}

Figure~\ref{Rs(P)} shows the \emph{mw} surface resistance of the \BSCCO\ wire as a function of the maximum value of the \emph{mw} magnetic field inside the cavity at the resonant frequencies of different TEM modes. As one can see, above a threshold value of $H_{mw}$, slightly depending on the frequency, $R_s$ deviates from its initial value, highlighting the occurrence of nonlinear effects. The \emph{mw} amplifier operating at $f\sim 2.5$~GHz allowed us to investigate a wide range of \emph{mw} fields; the largest value of $H_{mw}$ in the figure corresponds to an effective power level inside the cavity of the order of 37~dBm. At this frequency, nonlinear effects set in for $H_{mw} \gtrsim 40~\mu$T; the data above this threshold have been fitted by a $H_{mw}^n$ law, the continuous line in the figure shows the best-fit curve obtained with $n=0.63$. The inset is a plot of $\Delta R_s \equiv R_s(H_{mw})-R_s(0)$ as a function of $\Delta X_s \equiv X_s(H_{mw})-X_s(0)$, where $R_s(0)$ and $X_s(0)$ are the values of the \emph{mw} surface resistance and reactance, respectively, obtained at the lowest $H_{mw}$ value. The dashed line in the inset is the linear fit of the data of slope 1.37.

The nonlinear behavior of the \emph{mw} surface impedance in HTSC has been investigated by several authors especially in films~\cite{velichko,Oates_libro,karen}; only few investigations concern bulk samples~\cite{delayen,YBCO_Z(f)}. In spite of considerable effort, a universal model able to explain all the observed features has not yet developed. Reviews of the mechanisms responsible for the nonlinear \emph{mw} surface impedance have been reported in~\cite{samoilova,golo_NL,velichko,halbritter}. In granular samples, Josephson-fluxon motion in weak links plays an important role; in this case, $R_s(H_{mw})$ follows a $n$-power law, with $n$ ranging from 0.5 to 2, depending on the motion regime of Josephson fluxons~\cite{halbritter}. Our results suggest that the nonlinearity of \emph{mw} surface impedance in this sample arises from this mechanism. In this contest, from the expression reported by Halbritter~\cite{halbritter} for the flux flow-resistivity of Josephson fluxons, the results in the inset of Fig.~\ref{Rs(P)} give for the depinning frequency the value of $\approx 3.5$~GHz.

\section{Harmonic generation}
\subsection{Measurement technique}
In order to detect the power emitted by the \BSCCO\ wire at the SH and TH frequencies of the driving field it is necessary, at first, to properly move the two metallic screws to tune the first and second resonant modes of the cavity and/or the first and the third ones. To make these measurements we have used the following procedure. The continuous wave generated by the NA at the frequency of the fundamental mode is modulated to obtain a train of \emph{mw} pulses, with pulse width 5~$\mu$s and pulse repetition rate 100 Hz, and amplified up to a peak power level of $\approx$~44 dBm. The pulsed wave is filtered by a low-pass filter, reducing any harmonic content by more than 60 dB, and feeds the fundamental mode of the cavity through the excitation line. The wave coming from the cavity through the detection line, which contains components oscillating at harmonic frequencies because of the nonlinear response of the SC, is filtered by a band-pass filter, with more than 60 dB rejection at the fundamental frequency, and sent to a superheterodyne receiver. The latter is equipped by a 30 MHz logarithmic amplifier that provides an output voltage proportional to the harmonic power (noise level $\approx -75$~dBm). The signal is displayed by an oscilloscope and recorded on a computer by an IEEE-488 interface, which allows one to automatically acquire the experimental data.

\subsection{Results and discussion}
\begin{figure}[h]
\centering
\includegraphics[width=8cm]{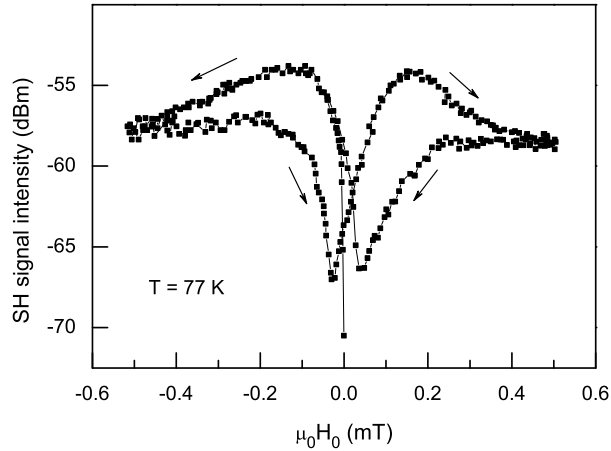}
\caption{SH signal intensity as a function of the DC magnetic field, $H_0$. Effective peak power in cavity at 2.5~GHz $\approx~20~\mathrm{dBm}$. The arrows indicate the field-sweep direction.}
\label{SH(77)}
\end{figure}

The SH signal intensity as a function of the DC magnetic field, $H_0$, is shown in Fig.~\ref{SH(77)}; the measurements have been performed at $T = 77$~K. The effective peak power in cavity at the fundamental mode, resonating at $f \sim 2.5$ GHz, is $\approx 20$~dBm. We have indicated as SH signal intensity the effective power at $2f$ inside the cavity, which has been calculated from the detected power properly rescaled by taking into account the coupling coefficient of the detection line. Since no SH emission is expected for \emph{mw} magnetic fields perpendicular to $\mathbf{H_0}$, the DC magnetic field has been oriented perpendicularly to the cylinder axis. By changing the orientation of $\mathbf{H_0}$, we have verified that the SH signal intensity significantly decreases when the component of $\mathbf{H_0}$ perpendicular to the wire axis decreases. However, a very weak SH signal is observed also for $\mathbf{H_0}$ parallel to the wire axis; this may be due to a little misalignment. The SH signal intensity shows a magnetic hysteresis having a butterfly-like shape, which has been already observed in inhomogeneous SC, whether at \emph{mw} frequencies~\cite{Noi-Metamat,Noi-JSUP,noiBKBO-SH} or at lower frequencies~\cite{Sen,Roy,Lam-Jef}.

Measurements performed sweeping $H_0$ in different ranges from $-H_{max}$ to $+H_{max}$ have shown that the hysteresis appears for $H_{max} \approx 0.2$~mT; on increasing $H_{max}$, the amplitude of the hysteresis loop increases and the sharp minima at low fields become less pronounced until they almost completely disappear for $H_{max}\gtrsim 2$~mT. Moreover, measurements performed at different input power levels have shown that the features of the SH vs $H_0$ curves do not exhibit noticeable variations.

\begin{figure}[h]
\centering
\includegraphics[width=8cm]{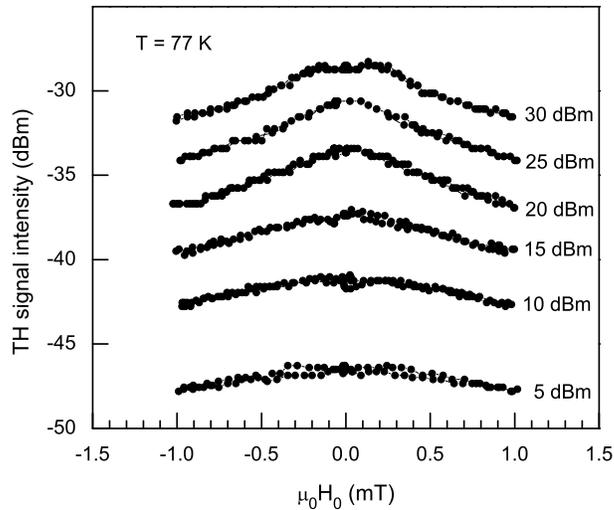}
\caption{TH signal intensity as a function of the DC magnetic field, at different input power levels (the values indicated on the right side of the plot refer to the effective peak power in cavity at the the resonant frequency of the fundamental mode).}
\label{TH(77)}
\end{figure}

Figure~\ref{TH(77)} shows the TH signal intensity as a function of the DC magnetic field, at different input power levels. In order to compare the SH and TH response in the same field geometry, $\mathbf{H_0}$ is perpendicular to the cylinder axis. By varying the DC field orientation, the features of the TH vs $H_0$ curves do not significantly change; only an increase of the signal intensity of about 1 dB is observed for $\mathbf{H_0}$ parallel to the wire axis. Measurements performed for different $H_{max}$ values have shown that the TH vs $H_0$ curve exhibits a monotonic decrease of the signal intensity with $H_0$ and no hysteresis is observed until $H_{max}$ reaches values larger than 2~mT.

By comparing the results of Figs.~\ref{SH(77)} and \ref{TH(77)}, one can note that the TH signal is much more intense than the SH one. As an example, the intensities of the SH and TH signals, generated with effective peak power at the fundamental mode of 20~dBm, differ of about 20 dB.

\begin{figure}[h]
\centering
\includegraphics[width=8cm]{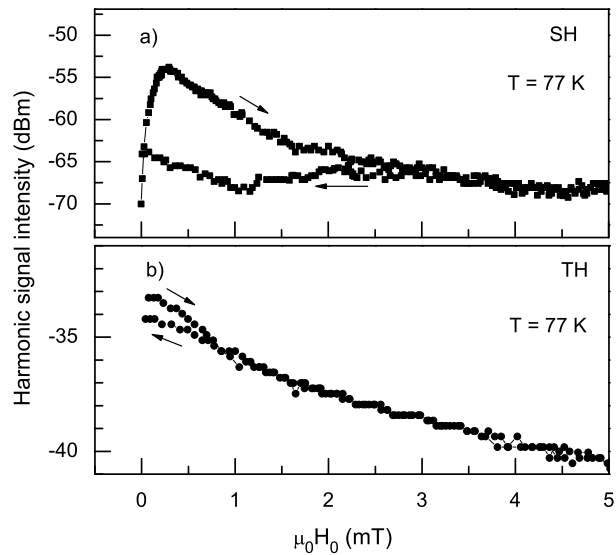}
\caption{a) SH- and  b) TH-signal intensity as a function of the DC magnetic field; input peak power $\approx 20$~dBm.}
\label{SH-TH}
\end{figure}

Figure~\ref{SH-TH} shows the field dependence of both SH and TH signal intensity, obtained at $T= 77$~K by sweeping $H_0$ from zero to 5~mT and back. In the SH signal, the hysteresis is enhanced, extends in a wide field range, and the sharp minima in the decreasing-field branch are no longer present. On the contrary, only a small hysteresis is observed in the TH signal, which extends in a narrow field range. The large difference in the SH and TH signal intensities is still present and enhanced (at high fields); this is so because the SH signal reduces faster with the field than the TH one. We would like to remark that it was possible to highlight this peculiarity of the nonlinear \emph{mw} emission just because we used the same device (with the same excitation conditions) to detect both the SH and TH emission.

Harmonic emission has been investigated in both conventional and high-$T_c$ SC; several mechanisms have been recognised, they play a different role depending on temperature, applied magnetic field, frequency and type of SC~\cite{samoilova,golo_NL,noi_review}. It is widely accepted that, at temperature not very near $T_c$, the mechanisms that play the most important role are i) nonlinear electromagnetic response of the weak-link matrix (at low magnetic fields) and ii) motion of Abrikosov vortices in the critical state (at $H > H_{c1g}$).

In SC containing weak links, essentially two mechanisms of nonlinearity may come into play. Harmonic emission is expected when supercurrents are induced by the DC and \emph{mw} magnetic fields in loops containing Josephson junctions. In this case, it is strictly related to the intrinsic nonlinearity of the Josephson current, $J_{cj}$~\cite{Noi_FisicaC161,Lam-Jef}. In addition, intergrain dynamics of Josephson fluxons may give rise to harmonic emission~\cite{Ji,muller,clem}. The results reported in Figs.~\ref{SH(77)} and \ref{TH(77)} should arise from these mechanisms; however, no hysteresis is expected by none of the models describing these processes.

Magnetic hysteresis in the harmonic signals, as well as in the AC susceptibility, has been detected in ceramic SC at low frequencies (100 Hz - 100 kHz)~\cite{Sen,Roy,Lam-Jef}, as well as at \emph{mw} frequencies~\cite{Noi-Metamat,Noi-JSUP}. At low frequencies, magnetic hysteresis in the same field range has been observed in the first, second and third harmonics, whose features strongly depend on the AC-driving-field intensity. The hysteretic behavior has been ascribed to the fact that, after the DC magnetic field has reached values large enough to allow Abrikosov fluxons penetrating the superconducting grains, the intergrain effective magnetic field in the decreasing-field branch is smaller than that in the increasing one in the virgin sample~\cite{Sen,Roy}, reducing $J_{cj}$. This process can justify the hysteresis we have detected in the harmonic response for $H_{max} > 2$~mT (see Fig.~\ref{SH-TH}), but it does not justify the hysteretic behavior of the SH signal at very low fields (well visible for $H_{max} \approx 0.2$~mT), too small for creating intragrain vortices. Furthermore, we have not observed a noticeable variation in the hysteresis shape at different input power levels.

In Refs.~\cite{Noi-JSUP,noiBKBO-SH}, we suggested that the hysteretic behavior of the \emph{mw} SH response in inhomogeneous SC can be related to the fact that both the above-mentioned nonlinear processes involving weak links come into play simultaneously. On the other hand, theoretical investigation performed by McDonald and Clem~\cite{clem} on the electromagnetic response of Josephson junctions has shown that SH signals can be generated, whose sign is different whether fluxons nucleate or exit (i.e. the phase depends on the magnetic-field-sweep direction). On the contrary, the phase of the even-harmonic signals arising from the nonlinearity of $J_{cj}$ is not expected to depend on the field-sweep direction, but only on the polarity of $H_0$ with respect to $H_{mw}$. So, the interference of the two signals may give rise to the hysteretic behavior when $H_{max} \geq H_{c1j}$. The finding that we do not observe hysteresis in the TH signal at very low fields confirms this scenario;  indeed, the phase of the odd-harmonic signals is expected to be independent of both the field-sweep direction and the $H_0$ polarity and no hysteresis comes out.

On the base of all the above considerations, our results suggest that in this wire $H_{c1j} \approx 0.2$~mT (the value of $H_{max}$ at which the hysteresis appears in the SH signal) and $H_{c1g} \approx 2$~mT (the value of $H_{max}$ at which the hysteresis appears in the TH signal). At higher DC fields, Abrikosov vortices penetrate the superconducting grains and additional mechanisms, due to dynamics of Abrikosov vortices, come into play, changing the features of both SH and TH vs $H_0$ curves. For $H_{max} > H_{c1g}$, the low-field signals in the decreasing-field branch reduces because some Josephson junctions are decoupled by the trapped flux.

Although the nonlinear processes occurring in weak links justify qualitatively the \emph{mw} harmonic emission, a model able to quantitatively account all the features of the harmonic signals has not yet been developed. Numeric calculations in the framework of the models elaborated to describe nonlinear emission by weak links~\cite{Lam-Jef,Ji,muller} have shown that the intensity of the SH and TH signal, at their maximum value, are of the same order of magnitude. Furthermore, the TH vs $H_0$ curve results structured, with maxima and minima at low fields, while we observe a monotonically decreasing with $H_0$. These results, and mostly the higher intensity of the TH signal, suggest that additional effects have to be taken into account to fully understand the nonlinear \emph{mw} emission; so, further investigation is necessary.

\section*{Conclusion}
We have built a tunable \emph{mw} coaxial cavity constituted by an outer Cu cylinder and an inner \BiSCCO\ wire. The cavity has been used to perform several kinds of measurements, useful for investigating the \emph{mw} response of the superconducting inner wire. In the present paper, we have reported results of the \emph{mw} surface resistance as a function of the frequency and the \emph{mw} magnetic field. Moreover, by tuning the different modes of the cavity to make them resonant at exactly harmonic frequencies, we have detected the power emitted by the inner wire at the second- and third-harmonic frequency of the driving field. The intensity of the second- and third-harmonic signal has been investigated as a function of an external DC magnetic field at relatively low fields. Our results have been discussed in the framework of the models reported in the literature for the electromagnetic response of granular superconductors. All the results have been qualitatively accounted for considering processes occurring in weak links; however, a quantitative agreement, specially for the third-harmonic signal, has not been obtained. We have shown that this type of device provides a useful tool for a complete investigation of superconductors of large dimension, both in the linear and nonlinear regimes. The advantage to use such device is to measure different \emph{mw} properties of the superconductor in the same excitation conditions. For example, the opportunity of investigating the second- and third-harmonic emission with the same device, allowed us to highlight that third-harmonic emission is much more enhanced than second-harmonic one. This disagrees with the results expected from the models reported in the literature for nonlinear effects at low fields, arising from weak links. We would like to remark that by this device one can perform further measurements: for example, both the \emph{mw} surface resistance and the harmonic signals can be measured as a function of the DC magnetic field at higher fields for investigating the Abrikosov-fluxon dynamics. Moreover, using a detection system working in a wider frequency range, higher harmonics can also be investigated.

\end{document}